\documentclass[aps,prb,showpacs,preprint,superscriptaddress,amsfonts,amssymb,amsmath]{revtex4}

\usepackage{amsfonts}
\usepackage{amsmath}
\usepackage{amssymb}
\usepackage{graphicx}

\usepackage{graphicx}
\begin{document}

\title{Bias-controllable intrinsic spin polarization in a quantum dot}
\author{Qing-feng Sun$^{1}$ and X. C. Xie$^{2,}$}
\affiliation{Beijing National Lab for Condensed Matter Physics and
Institute of Physics, Chinese Academy of Sciences, Beijing
100080, China \\
$^2$Department of Physics, Oklahoma State University, Stillwater,
Oklahoma 74078 }
\date{\today}

\begin{abstract}
We propose a novel scheme to efficiently polarize and manipulate
the electron spin in a quantum dot. This scheme is based on
the spin-orbit interaction and it possesses following advantages:
(1) The direction and the strength of the spin polarization is
well controllable and manipulatable by simply varying the bias or
the gate voltage. (2) The spin polarization is quite large even
with a weak spin-orbit interaction. (3) Both electron-electron
interaction and multi-energy levels do not weaken but strengthen
the spin polarization. (4) It has the short spin flip time. (5)
The device is free of a magnetic field or a ferromagnetic
material. (6) It can be easily realized with present technology.
\end{abstract}


\maketitle

How to efficiently control and manipulate the spin is an important
and challenging issue in spintronics.\cite{ref1,ref2}
Normally, the spin is difficult to be manipulated by a voltage-bias because
the bias (or an electric field) does not act on the spin.
Alternative methods, such as using a magnetic
field or polarized light to manipulate the spin have been
suggested, but they are far from in real use.

The quantum dot (QD) is an elementary cell of nano-electronic
devices. The electron spin in the QD has been suggested as an ideal
candidate for the qubit. Electron spin automatically
comprises two levels that is a natural representation of a
qubit,\cite{addref1,ref3,ref4} moreover the spin has a long
decoherent time. However, in order to utilize the electron spin
in the QD as a qubit, one first has to figure out how to efficiently
polarize and manipulate the spin in the QD, i.e. writing a
spin into the QD. One natural idea is to couple the QD to a
ferromagnetic (FM) lead, such that the polarized spin in the FM
can be injected into the QD.\cite{addref1,ref5,addref2} Another
idea is to use an external magnetic field to polarize the spin in
the QD.\cite{ref3} But both methods are not feasible in current
experiments, because first it is very difficult to inject the spin
from a FM into a semiconductor,\cite{ref6} and for the second
proposal to succeed, one needs a very strong external magnetic
field confined to a small region of a QD.

Recently, based on the spin-orbit (SO) interaction, several
theoretical studies have proposed that spontaneous spin
accumulation can take place. For example, in the confined spin
Hall devices, the opposite spin accumulations emerge at the
boundaries of the samples.\cite{ref7,ref8,ref9,ref10} Can one
achieve an effective spin manipulation in a QD by using the SO
interaction?

In this Letter, we propose a new scheme to polarize and manipulate
the spin in a QD by using the SO interaction. The main idea is as
follows. Consider a QD coupled to two (left and right) leads and
there also exists a direct bridge coupling between the two leads
(see Fig.1a). In this system, an electron from the QD tunnelling
to the left lead or vice versa has two paths: one path is through
the direct tunnelling, the other is for the electron to first
travel to the right lead and follow up with a tunnelling to the
left lead through the bridge coupling (see Fig.1a). Let $t_{ij}$
describes the transmission coefficient from $j$ to $i$, with $i,j
=$ $L$(left lead), $R$ (right lead), and $d$ (QD). Assume that the
QD or the bridge arm contains the Rashba SO
interaction,\cite{ref11,ref12} a spin-dependent extra phase
$\sigma\varphi$ is generated in the path, thus, $t_{Rd}$ changes
into $t_{Rd}e^{i\sigma\varphi}$ (the phase $\varphi= -k_R L
=-\alpha_{R} m^* L/\hbar^2$ also describes the spin precession
angle, with $\alpha_R$ being the Rashba SO interaction constant
and $L$ being the size of the QD).\cite{ref13,ref14} Then the
total effective coupling (or tunnelling) strength $T_{L\sigma}$
between the QD and the left lead is:
\begin{eqnarray}
 T_{L\sigma} & =  & |t_{Ld} +t_{LR}(-i\pi \rho) t_{Rd}
 e^{i\sigma\varphi}|^2 \nonumber \\
 & =&
   |t_{Ld}|^2 +|\pi \rho t_{LR}t_{Rd}|^2
    +2\pi\rho |\bar{t}| \sin(\phi_0 +\sigma\varphi),
\end{eqnarray}
where $\bar{t}=t_{LR} t_{Rd} t_{Ld}^*$, $\phi_0$ is the phase of
$\bar{t}$, and $\rho$ is the density of states in the lead.
Similarly, the total effective coupling strength $T_{R\sigma}$ for
the QD and the right lead is:
\begin{eqnarray}
 T_{R\sigma} & =  & |t_{Rd} e^{i\sigma\varphi}
                    + t_{LR}^*(-i\pi \rho) t_{Ld} |^2 \nonumber \\
 & =&
   |t_{Rd}|^2 +|\pi \rho t_{LR}t_{Ld}|^2
    -2\pi\rho |\bar{t}| \sin(\phi_0 +\sigma\varphi).
\end{eqnarray}
In general $T_{\alpha\uparrow}$ ($\alpha=L,R$) is different from
$T_{\alpha\downarrow}$. If $T_{L\uparrow}>T_{L\downarrow}$, [then
$T_{R\uparrow}$ must be $< T_{R\downarrow}$ from Eqs.(1,2)], it is
easier for the spin-up electron to tunnel from the left lead into
the QD than for the spin-down electron, but it is more difficult
for it to tunnel out from the QD to the right lead because
$T_{R\uparrow} < T_{R\downarrow}$. Therefore the QD should be spin
polarization in the `up' (or `down') direction when the left lead
is the high (or low) voltage terminal. Our following detailed
numerical investigation indeed shows that the QD is spin polarized
under the non-zero bias. The spin polarization is quite large even
with a weak SO interaction and in a small QD. Particularly, the
strength and the direction of the spin polarization are easily
controllable and manipulatable by varying the bias or the gate
voltage.

Our device is described by the following Hamiltonian:\cite{ref14}
\begin{eqnarray}
H &=& \sum\limits_{k,\sigma,\alpha }
       \epsilon_{\alpha k} a^{\dagger}_{\alpha k \sigma} a_{\alpha k \sigma}
     + \sum\limits_{\sigma} \epsilon_d d^{\dagger}_{\sigma}
          d_{\sigma}
      + U d^{\dagger}_{\uparrow}d_{\uparrow}
         d^{\dagger}_{\downarrow}d_{\downarrow} \nonumber\\
  & + & \sum\limits_{k,k',\sigma} t_{LR} \left[
        a^{\dagger}_{Lk\sigma} a_{Rk'\sigma}    + a^{\dagger}_{Rk'\sigma} a_{Lk\sigma}
        \right] \nonumber \\
  & + & \sum\limits_{k,\sigma} \left[
       t_{Ld} a^{\dagger}_{Lk\sigma} d_{\sigma} +
        t_{Rd} e^{ i \sigma \varphi} a^{\dagger}_{Rk\sigma} d_{\sigma}
\right]
        +H.c.
\end{eqnarray}
where $d_{\sigma}$ and $a_{\alpha k\sigma}$ are annihilation
operators in the QD and the lead $\alpha$, respectively. The QD
consists a single energy level and an electron-electron (e-e)
interaction $U$. Consider there exists the Rashba SO interaction
in the QD, an extra phase $i\sigma\varphi$ is added in the hopping
term of $t_{Rd}$.\cite{ref14} We emphasize that the system
contains no magnetic field and the sample is not a FM material.

The intradot spin-up (or spin-down) electronic occupation number
$n_{\sigma}$ can be solved by using the standard Keldysh
nonequilibrium Green's function method. Following the procedure of
our previous paper,\cite{ref14} the retarded Green function
$G^r_{d\alpha\sigma}$ is obtained as:
\begin{eqnarray}
  G^r_{d\alpha\sigma}(\omega) & = &
     G^r_{dd\sigma} (\tilde{t}_{d\alpha\sigma} + \tilde{t}_{d\bar{\alpha}\sigma}
      g^r_{\bar{\alpha}}
    t_{\bar{\alpha}\alpha} ) g^r_{\alpha} /A ,
\end{eqnarray}
where $ G^r_{dd\sigma} (\omega)  = 1/ \{g^{r-1}_{d\sigma}
-\sum_{\alpha} (\tilde{t}_{d\alpha\sigma} +
\tilde{t}_{d\bar{\alpha}\sigma} g^r_{\bar{\alpha}}
t_{\bar{\alpha}\alpha} ) g^r_{\alpha} \tilde{t}_{\alpha d
\sigma}/A \}$, $A=1-g^r_R t_{RL} g^r_L t_{LR}$,
$\tilde{t}_{Ld\sigma}=\tilde{t}^*_{dL\sigma} =t_{Ld}$,
$\tilde{t}_{Rd\sigma} = \tilde{t}_{dR\sigma}^* = t_{Rd}
e^{i\sigma\varphi}$, and $\bar{\alpha} = R$ for $\alpha =L$ or
$\bar{\alpha} = L$ for $\alpha =R$. The Green functions $g^r$ are
for the decoupled system (i.e. when $t_{LR}=t_{Ld}=t_{Rd}=0$),
with $g_{d\sigma}^r(\omega)=
\frac{\omega-\epsilon_{d}-U+Un_{\bar{\sigma}}
}{(\omega-\epsilon_{d})(\omega-\epsilon_{d}-U) }$ and
$g^r_R(\omega)=g^r_L(\omega) = -i\pi \rho$. Then the occupation
numbers are: $ n_{\sigma} = -i \int \frac{d\omega}{2\pi}
G^<_{dd\sigma}(\omega)$, and $G^<_{dd\sigma}(\omega) =
\sum_{\alpha} |G^r_{d\alpha \sigma}(\omega)|^2 2i
f_{\alpha}(\omega)/(\pi\rho)$, where $f_{\alpha}(\omega) =
[exp((\omega-\mu_{\alpha})/k_BT)+1]^{-1}$ is the Fermi
distribution function in the lead $\alpha$.

Next we present our numerical investigation. In all numerical
calculations, we take $\rho=1$ and symmetric coupling strengths
$t_{Ld}=t_{Rd}=0.4$ (the corresponding line-width $\Gamma =2\pi
\rho |t_{L(R)d}|^2 \approx 1$). The chemical potential $\mu_L =
-\mu_R =V/2$ with the bias $V$. Fig.2 shows the occupation number
$n_{\sigma}$ and the spin accumulation $\Delta n$ ($\Delta n
\equiv n_{\uparrow} - n_{\downarrow}$) versus the intradot level
$\epsilon_d$ (Fig.2a-d) and the bias $V$ (Fig.2e-h). The non-zero
$\Delta n$, i.e. the spin polarization in the QD, indeed emerges
under a finite bias. $\Delta n$ has the following features: (1)
When the bias $V=0$, $\Delta n$ is identically zero for any
$\epsilon_d$ because of the time reversal invariance.\cite{ref14}
(2) With the bias $V$ increasing from 0, $\Delta n$ increases.
While $V/2
>|\epsilon_d|$, i.e. $\mu_L >\epsilon_d>\mu_R$, $\Delta n$ is already
large and the QD is well spin polarization (see Fig.2f,h). If the
bias is reversed, $\Delta n$ changes its sign, i.e. the spin
polarized direction is reversed. This means that the direction and
the strength of the spin polarization are easily controlled and
tuned by changing of the external bias. (3) For a fixed bias $V$
with varying $\epsilon_d$ by tuning the gate voltage, $\Delta n$
can also be modulated (see Fig.2b,d). When $\epsilon_d$ is above
both $\mu_L$ and $\mu_R$, $n_{\uparrow}$ and $n_{\downarrow}$ are
almost zero. On the other hand, if $\epsilon_d < \mu_L, \mu_R$,
$n_{\uparrow},n_{\downarrow} \approx 1$. But when the energy level
$\epsilon_d$ is in the bias window with $\mu_L
>\epsilon_d
>\mu_R$, $\Delta n$ is quite big and the QD is largely spin
polarized. (4) Even for a small $\varphi$, $\Delta n$ is large.
For example, $\varphi =0.2$, $\Delta n$ is near 0.2 (see
Fig.2b,f). While $\varphi =\pi/4$, $\Delta n$ can be over 0.5 (see
Fig.2d,h), which is quite large for spin polarization.

In Fig.3a,b we show $\Delta n$ dependence on the phase $\varphi$
and the bridge coupling strength $t_{LR}$. $\Delta n$ versus
$\varphi$ exhibits a periodic function with the period of $2\pi$.
While $\varphi = \pm \pi/2$, $\Delta n$ is near $\pm 1$ and the
spin polarization can reach almost $100\%$. $\Delta n$ versus
$t_{LR}$ is shown in Fig.3b, in which $\Delta n =0$ at $t_{LR}=0$
because of the shut-down of the bridge coupling. With a gradual
opening of the bridge coupling (i.e. the gradual raising of
$t_{LR}$), $\Delta n$ increases first and follows by a slight
reduction. But $\Delta n$ still is over $0.1$ even at quite large
values of $t_{LR}$ (with $\varphi=0.2$).

Following, we investigate the effect of the e-e interaction (i.e.
$U \not=0$), which is shown in Fig.4a-d. In general, the
interaction $U$ increases $\Delta n$ because of the repulsive
interaction between two electrons, namely strengthens the spin
polarization. In particular, it has the following features: (1)
For a wide range of $\epsilon_d$, $\Delta n$ can maintain large
values (see Fig.4b). In fact, $\Delta n$ is large as soon as
$\epsilon_d$ or $\epsilon_d +U$ is within the bias window. (2)
$\Delta n$ is larger than the value with $U=0$. For example for
the case of $\varphi=0.2$, $\Delta n$ only reaches $0.18$ at $U=0$
(see Fig.2b,f). However, at $U=10$, $\Delta n$ is about $0.24$ for
a large range of $\epsilon_d$. Furthermore, $\Delta n$ can reach
up to $0.33$ at some special values of $\epsilon_d$ (see Fig.4b).
Correspondingly, the spin polarization $p$ [$p\equiv \Delta
n/(n_{\uparrow}+n_{\downarrow})$] can reach $30\%$ for that range
of $\epsilon_d$ and $42\%$ at those special values of
$\epsilon_d$. Note this spin polarization is indeed fairly large,
although $\varphi$ is only $0.2$. (3) With the bias $V$ increasing
from 0, $\Delta n$ can quickly increase as shown in Fig.4d. While
$U=0$, $\Delta n$ reaches $0.17$ until $V=5$. However, when $U
\not=0$ (e.g. $U=3$ or $5$), $\Delta n$ has exceeded $0.18$ at
$V=1$ (see Fig.4d).

In the above model [or the Hamiltonian (3)], only one energy level
in the QD is considered. How is the spin accumulation $\Delta n$
affected by the multi-levels in the QD? In fact, if there is only
one level in the bias window and the others are outside the bias
window, the outside levels do not affect the spin accumulation,
the system acts as if it is a one-level system. On the other hand,
if there are $N$ ($N>1$) energy levels in the bias window, then
each level will contribute a $\Delta n$ because the mechanism
mentioned in the introduction [see Eq.(1,2)] is effective for each
level. So the total spin accumulation $\Delta n_T$ is
approximatively $N \Delta n$ and is strongly enhanced.

How is time required for the spin flip to take place under a
reversed bias? In other words, to consider that the bias is
positive $V$ in the time $t<0$ (so the QD has the spin
polarization in $+z$ direction with a positive $\Delta n$), and
the bias is reversed at $t=0$ and it keeps the value $-V$ all
along in the time $t>0$. After this bias reversal, how long does
it take for $\Delta n$ to change its sign? In order to answer this
question, we have to solve the time-dependent occupation number
$n_{\sigma}(t)$. From the Keldysh equation, we have:
\begin{eqnarray}
  n_{\sigma}(t) & = &  -i G^<_{dd\sigma}(t,t) \nonumber\\
  &  = &
   \sum\limits_{\alpha} \iiiint dt_1 dt_2 dt_3 dt_4
   G^r_{d\alpha\sigma}(t,t_1)g^{r-1}_{\alpha}(t_1,t_2)
   g^<_{\alpha}(t_2,t_3) g^{a-1}_{\alpha}(t_3,t_4)
   G^a_{\alpha d\sigma}(t_4,t).
\end{eqnarray}
In the present case, although the bias is reversed at time $t=0$,
the retarded (advanced) Green functions $G^{r(a)}(t,t_1)$ and
$g^{r(a)}(t_1,t_2)$ are not affected (at $U=0$), and they are
still functions of the time difference. For example,
$G^r_{d\alpha\sigma}(t,t_1) =\int \frac{d\omega}{2\pi}
e^{-i\omega(t-t_1)} G^r_{d\alpha\sigma} (\omega)$ and
$G^r_{d\alpha\sigma}(\omega)$ are identical with the ones in
Eq.(4) that are for a constant-biased case. The Keldysh Green
function $g^<_{L/R}(t_1,t_2)$ for the decoupled lead in Eq.(5) is:
$
 g^<_{L/R}(t_1,t_2)
  =  i\rho \int d\omega f(\omega) e^{-i\omega(t_1-t_2)}
 e^{i (+/-) (|t_1|-|t_2|)V/2}
$, with $f(\omega) =1/\{exp(\omega/k_BT)+1\}$.

The numerical results of $n_{\sigma}(t)$ and $\Delta
n(t)=n_{\uparrow}(t)-n_{\downarrow}(t)$ versus the time $t$ are
shown in Fig.5. $\Delta n(t)$ shows a quick reversal when the bias
is reversed. For example, for the parameters of Fig.5a, $\Delta
n(t) \approx 0.564$ for $t \leq 0$. When the bias is reversed at
$t=0$, $\Delta n(t)$ quickly reverses. When $t=3/\Gamma$, $\Delta
n(t) \approx -0.549$, thus, well reversed. If to take $\Gamma =
0.1 meV$,\cite{ref15} the reversal time $t =3/\Gamma \approx
2\times10^{-11} s$, that is very short.

Before summary, we discuss the realizability. We suggest a
possible experimental setup as shown in Fig.1b. The device is
fabricated with a two-dimensional electron gas (e.g. the one in
Ref.\cite{ref16}). The dark region is the etching region or the
deposited metal split gate with applied negative voltage to
control the coupling coefficients $t_{LR}$, $t_{Ld}$, and
$t_{Rd}$. The electrons are not present in the dark region and a
QD is formed in the lower arm. A gate voltage $V_g$ is applied
above the QD to control the Rashba SO interaction constant
$\alpha_R$ as well the intradot energy levels. This device with
its size within the phase coherent length can be easily realized
with today's semiconductor technology.\cite{ref15,ref16} The
parameters of the bias $V$, the coupling coefficients $t_{LR}$,
$t_{Ld}$, and $t_{Rd}$, can be conveniently tuned to satisfy the
condition for substantial spin polarization. Next we discuss the
phase parameter $\varphi$ (i.e. the spin precession angle) and the
temperature effects. In our proposed scheme, even for quite small
$\varphi$ (e.g. $0.2$), $\Delta n$ is already large. In a recent
experiment,\cite{ref16} $\varphi$ was successfully modulated over
$0.75\pi$ with size $L=1.5\mu m$ (correspondingly $\alpha_R
\approx 2\times 10^{-12} eVm$). If the size of our QD is $200nm$,
$\varphi$ should be tunable in the range $0.1\pi \approx 0.3$.
Moreover, some experiments have measured that $\alpha_R$ can reach
$3\times 10^{-11} eVm$,\cite{refnn} then $\varphi =0.2$ for a QD
as small as $L \approx 10 nm$. So the parameter of $\varphi =0.2$
can be realized.\cite{note1} The temperature $k_BT$ is not a
problem with this scheme. Even with $k_BT /e = V/2$, the results
is almost unchanged. If one takes the bias $V=2mV$, $T =eV/2k_B
\approx 10K$. Finally, we compare this proposal to some recent
works on the spin Hall effect\cite{ref7,ref8,ref9,ref10}. The
opposite spin accumulations emerge at two opposite boundaries in a
confined spin Hall system. In contrast with those works, the size
of the present system is $L = \frac{\varphi}{2\pi}L_{SO}
=\frac{0.2}{2\pi}L_{SO} \approx 0.03L_{SO}$ ($L_{SO}\equiv 2\pi
\hbar^2 /\alpha_R m^* $) and this size is much smaller than the
confined spin Hall system for which the size is usually several
times of $L_{SO}$.

In summary, we have proposed a new method to generate the spin
polarized electrons in a quantum dot by utilizing the spin-orbit
(SO) interaction. A large spin polarization can be produced even
with a weak SO interaction and in a small dot. In particular, the
direction and the strength of the spin polarization can be
controlled and tuned by varying the bias or the gate voltage.

{\bf Acknowledgments:} We gratefully acknowledge financial support
from the Chinese Academy of Sciences and NSFC under Grant No.
90303016 and No. 10474125. XCX is supported by US-DOE under Grant
No. DE-FG02-04ER46124 and NSF-MRSEC under DMR-0080054.

\newpage
\begin{figure}

\caption{(Color online) (a) Schematic diagram for the system of
two leads coupled to a QD as well a bridge coupling between two
leads. (b) Schematic diagram for our proposed experimental device
fabricated in a 2DEGs. The dark regions are the split gate to
control the coupling coefficients $t_{LR}$, $t_{Ld}$, and
$t_{Rd}$. The inclined lattice region is the gate that controls
the Rashba SO interaction constant $\alpha_R$ and the level
$\epsilon_d$.}\label{fig:1}

\caption{(Color online) (a-d) are $n_{\uparrow}$ [solid curves in
(a) and (c)], $n_{\downarrow}$ [dotted curves in (a) and (c)], and
$\Delta n$ [in (b) and (d)] vs the level $\epsilon_d$ for the bias
$V=2$, $4$, and $8$ along the arrow direction. (e-h) are
$n_{\uparrow}$ [solid curves in (e) and (g)], $n_{\downarrow}$
[dotted curves in (e) and (g)], and $\Delta n$ [in (f) and (h)] vs
the bias $V$, where in (e) and (g) $\epsilon_d=-1$, 0, 1, and 2
from top to bottom, in (f) and (h) $\epsilon_d = 0$, 1, $-1$, and
$2$ along the arrow direction. Notice that in (f) the curve of
$\epsilon_d =0$ ($-1$) almost overlaps with one of $\epsilon_d= 1$
($2$) so that they cannot be seen in the figure. The other
parameters are: $t_{LR}=0.3$ and $\varphi=0.2$ [in (a), (b), (e),
and (f)], $t_{LR}=0.2$ and $\varphi=\pi/4$ [in (c), (d), (g), and
(h)]. The temperature $k_BT =0$ and $U=0$. }\label{fig:2}

\caption{(Color online) (a) $\Delta n$ vs $\varphi$ for
$t_{LR}=0.3$ and (b) $\Delta n$ vs $t_{LR}$ for $\varphi=0.2$. The
parameters are $\epsilon_d=U=k_BT=0$. }\label{fig:3}

\caption{(Color online) (a) and (b) are $n_{\uparrow}$ [the thick
curves in (a)], $n_{\downarrow}$ [the thin curves in (a)], and
$\Delta n$ vs $\epsilon_d$ for the bias $V=6$, and $U=3$ (the
dotted curves) and $10$ (the solid curves). (c) and (d) are
$n_{\uparrow}$ [the thick curves in (c)], $n_{\downarrow}$ [the
thin curves in (c)], and $\Delta n$ vs the bias $V$ for
$\epsilon_d=0$. The other parameters are $t_{LR}=0.3$,
$\varphi=0.2$, and $k_BT =0$. }\label{fig:4}

\caption{(Color online) $n_{\uparrow}$, $n_{\downarrow}$, and
$\Delta n$ vs the time $t$ when the bias $V$ is reversed at $t=0$,
where the parameters are: $\epsilon_d = U =k_BT =0$, $V=8$ in
$t<0$ and $V=-8$ in $t>0$. In (a) $t_{LR}=0.2$ and
$\varphi=\pi/4$, and in (b) $t_{LR}=0.3$ and $\varphi=0.2$.
}\label{fig:5}

\end{figure}


\begin{references}
\bibitem{note1}
In order to formed the QD, some split gates are added. Those split
gates perhaps affect the potential in the vertical direction of
2DEGs, so it maybe reduces or destroy the Rashba SO interaction in
the QD and then $\varphi$. We emphasize that it is nothing even
the Rashba SO interaction in the QD is completely destroyed.
Because that in our scheme it only need the phase (i.e. spin
precession) different for the electron travelling the QD and the
bridge arm. So even the phase $\varphi_{QD}$ of passing the QD is
zero, the spin polarization still emerges and the proposed scheme
can work as long as the existence of the phase in the bridge arm.
\end{references}
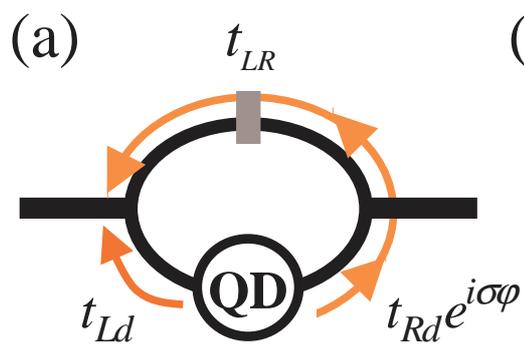
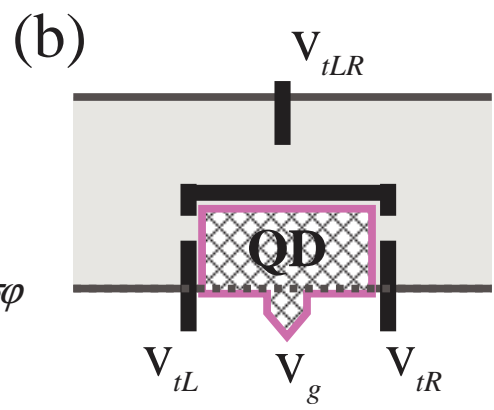
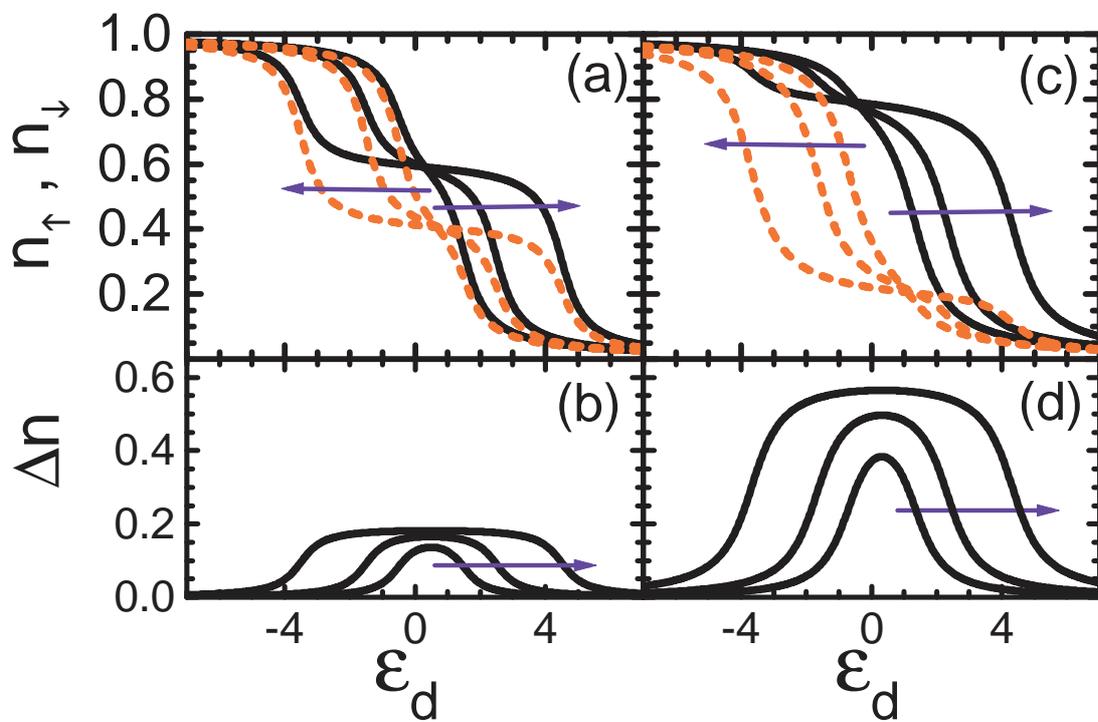
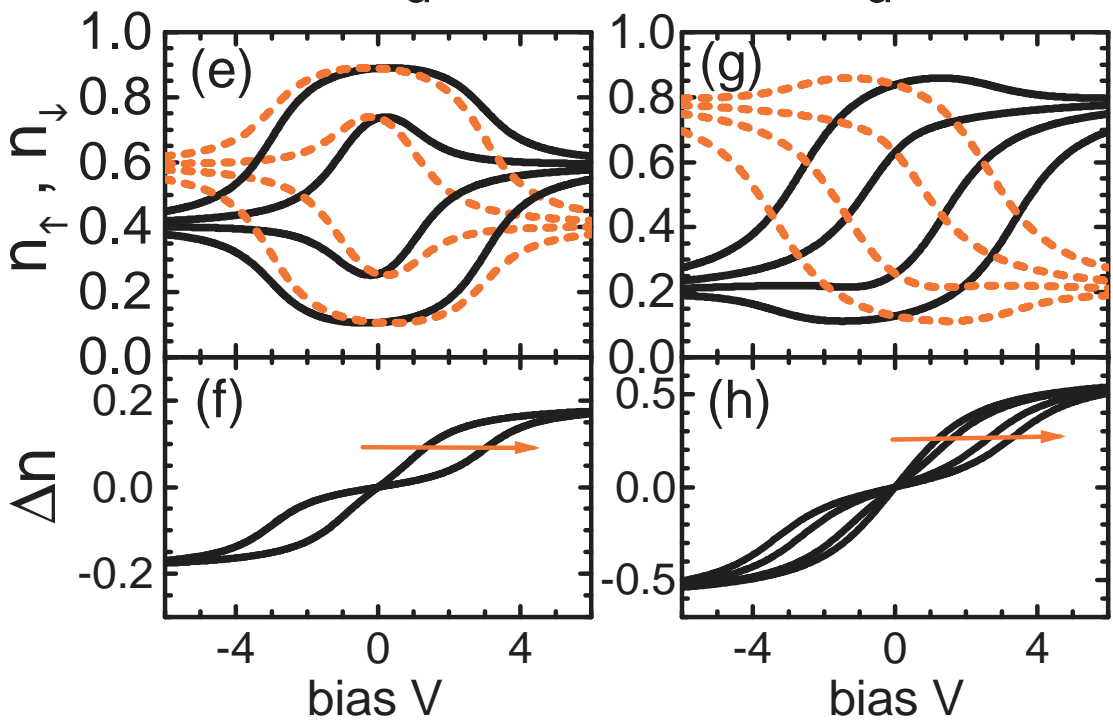
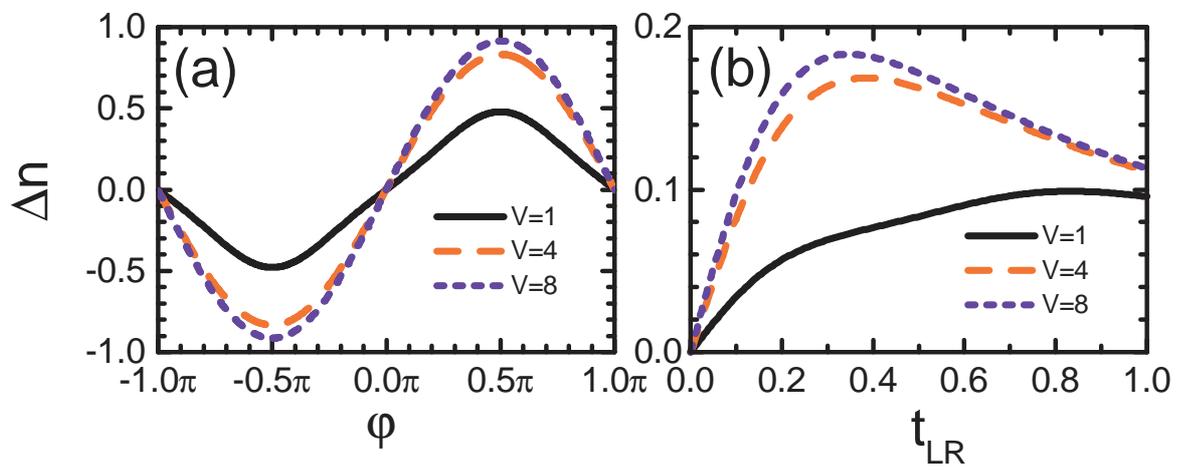
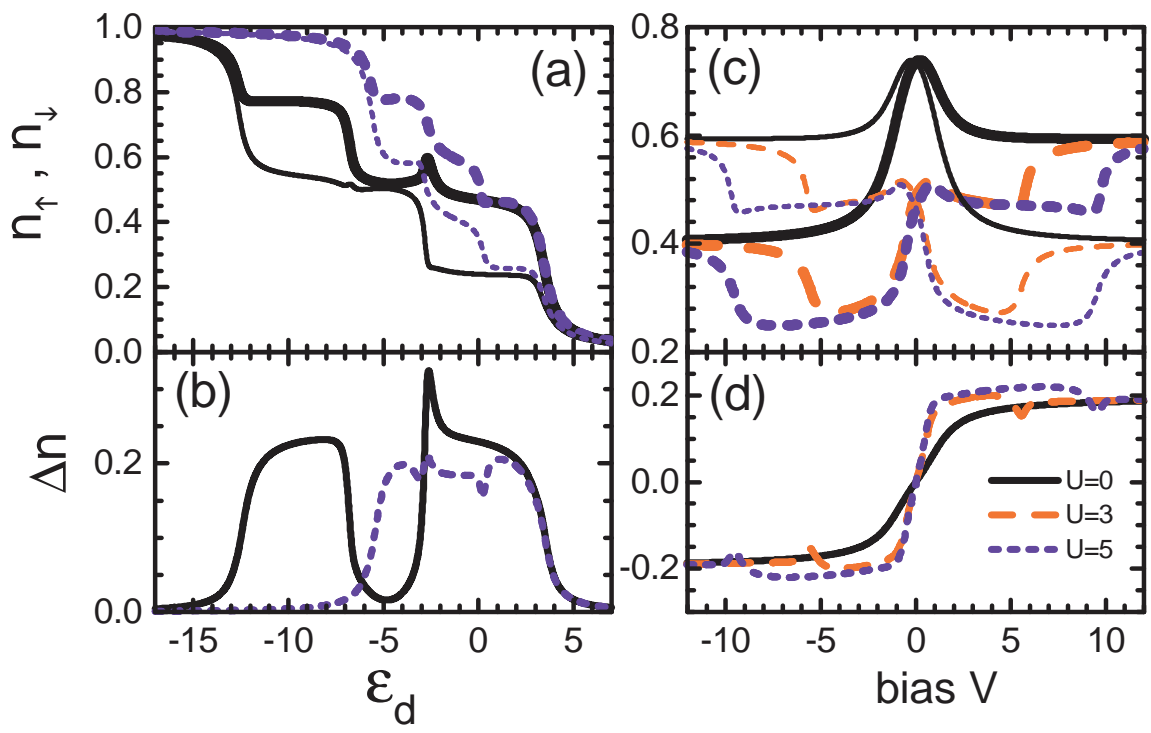
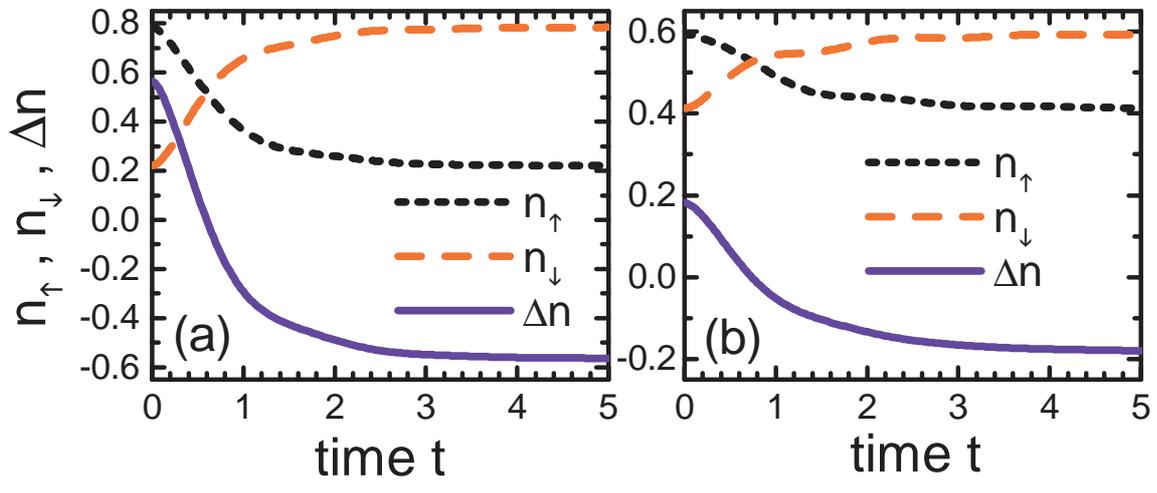
\end{document}